\documentclass[aps,prl,preprint,superscriptaddress,showpacs,showkeys,bibnotes,footinbib]{revtex4}
\usepackage{graphicx}
\usepackage{color}

\usepackage{epstopdf}
\newcommand{\comment}[1]{}
\newcommand{\silica}{${\rm SiO}_{2}$}
\newcommand{\tetra}{${\rm SiO}_{4}$}
\newcommand{\tetrahedron}{${\rm Si}({\rm O}_{4})_{1/2}$}

\newcommand{\IZA}[1]{\textbf{#1}}
\newcommand{\mat}[1]{\mbox{\boldmath{$\rm #1$}}}
\newcommand{\vctr}[1]{\mbox{\boldmath{$#1$}}}

\begin{document}

\title{Density of mechanisms within the flexibility window of zeolites}

\author{V.~Kapko}
\affiliation{Department of Physics, Arizona State University, Tempe, Arizona 85287, USA}

\author{C.~Dawson}
\affiliation{Department of Physics, Arizona State University, Tempe, Arizona 85287, USA}

\author{I.~Rivin}
\affiliation{Temple University, Department of Mathematics, Philadelphia, PA 19122, USA}

\author{M.~M.~J.~Treacy}
\email{treacy@asu.edu}
\affiliation{Department of Physics, Arizona State University, Tempe, Arizona 85287, USA}

\date{\today}

\begin{abstract}
By treating idealized zeolite frameworks as periodic mechanical trusses, we show that the number of flexible folding mechanisms in zeolite frameworks is strongly peaked at the minimum density end of their flexibility window. 25 of the 197 known zeolite frameworks exhibit an extensive flexibility, where the number of unique mechanisms increases linearly with the volume when long wavelength mechanisms are included. Extensively flexible frameworks therefore have a maximum in configurational entropy, as large crystals, at their lowest density. Most real zeolites do not exhibit extensive flexibility, suggesting that surface and edge mechanisms are important, likely during the nucleation and growth stage. The prevalence of flexibility in real zeolites suggests that, in addition to low framework energy, it is an important criterion when searching large databases of hypothetical zeolites for potentially useful realizable structures.
\end{abstract}


\maketitle

Zeolites are crystalline aluminosilicate materials that are microporous to small molecules \cite{Barrer78}. Because of their importance in the petrochemical and fine-chemical industries as catalysts and molecular sieves, there are ongoing efforts to synthesize new zeolite materials \cite{Degnan07}. At present there are 197 recognized zeolite framework types \cite{Atlas}, and that number has been doubling every 14 years. Recently, computer-generated databases of hypothetical zeolite frameworks have appeared with many millions of topologies with low calculated framework energies \cite{Treacy97,Treacy04,Deem09}. Given this abundance of energetically plausible structures, it has been puzzling why only a small subset of these structures actually occur in nature.

Real zeolites have complex structures and compositions. However, to explore their flexibility they can be idealized as tetrahedrally-coordinated \silica\ frameworks comprising perfect \tetrahedron\ tetrahedra. Oxygen atoms are shared between tetrahedra and form the linkages. In real zeolites, tetrahedra are indeed close to perfectly regular. The intertetrahedral Si-O-Si bond angle is weakly constrained near 145$^{\circ}$, but with a wide distribution of allowable angles. Dove and colleagues \cite{Dove93,Giddy93} showed that if tetrahedral silicates are modeled with force-free spherical joints at the oxygens, zero-frequency rigid-unit modes (RUMS) arise. Sartbaeva et al.~\cite{Sartbaeva06} showed that many RUM modes correspond to finite folding mechanisms, allowing the stress-free framework to explore a range of densities termed the flexibility window.

Remarkably, almost all of the known aluminosilicate zeolites exhibit a flexibility window \cite{Kapko10}, and the structures of real zeolites adopt the low density end of the window \cite{Sartbaeva06,Kapko10}. This state minimizes the framework electrostatic energy, largely by repulsion between oxygen anions \cite{OKeeffe77}. However, fewer than 10\% of the low energy hypothetical zeolites in one large database \cite{Treacy04} are flexible when represented as pure silicates.

Tetrahedral silicate frameworks are locally isostatic, since the six of degrees of freedom per tetrahedron are exactly matched by the six constraints on the oxygens (three constraints per oxygen, which are shared by two tetrahedra). The flexibility in zeolites arises mainly from their high crystallographic symmetry, which imposes redundant constraints; in terms of rigidity theory, zeolites have non-generic structures \cite{Giddy93, Hammonds97}. Periodicity adds six unit cell degrees of freedom \cite{Guest03}. Three of these correspond to rigid-body translations. The remaining three, however, do not correspond to rigid-body rotations, as such motions are incommensurate with the lattice \cite{Guest03}. These correspond to three internal flexing degrees of freedom. Consequently, all periodic frameworks in three dimensions are guaranteed at least three infinitesimal folding mechanisms \cite{Guest03, Kapko09}, regardless of additional crystallographic symmetry. These correspond to the three acoustic modes \cite{Hammonds98}. Infinite aperiodic frameworks are not guaranteed these mechanisms. An account of flexibility of periodic frameworks with internal symmetry has been given recently by Malestein and Theran \cite{Malestein11} and Ross et al.~\cite{Ross11}.

To explore the configurational space of the flexibility window we employ techniques from rigidity theory by regarding a zeolite framework as a periodic truss. The O-O edges of tetrahedra are treated as bars of ideal length $L\!=\!0.263$ nm, and the six-coordinated oxygen atoms behave as force-free spherical joints \cite{Pellegrino86,Guest03}. Si atoms are `carried' inside the rigid tetrahedra and are ignored. The compatibility matrix, \mat{C}, describes the $3N$ rigid constraints that maintain the bar lengths with $N$ oxygen atoms per unit cell. The vector of infinitesimal displacements of the oxygen atoms, $\mat{d}$, is related to the vector of O-O edge extensions, $\mat{e}$, via
\begin{equation}
\mat{C}\mat{d} = \mat{e}.
\label{eq:R}
\end{equation}
The condition $\mat{e}\!=\!0$ describes a rigid framework. \mat{C} is a $3N \times 3N$ matrix whose elements are $\tilde{x}_{ij}=(x_{j}-x_{i})/L$, $y_{ij}$, $z_{ij}$ etc, with appropriate adjustments for periodic continuity. If we admit unit cell degrees of freedom, \mat{d} acquires six additional components representing small changes in the unit cell dimensions, $\delta a$, $\delta b$, $\delta c$, $\delta \alpha$, $\delta \beta$ and $\delta \gamma$ \cite{Guest03}. \mat{C} is now a $3N \times (3N+6)$ matrix embodying $3N+6$ degrees of freedom and $3N$ constraints. With compatible initial coordinates that satisfy $\mat{e}=0$, $\mat{C}\mat{d}\!=\!0$ can be interpreted as an eigenvector problem with eigenvalues equal to zero. $\mat{C}$ is rank deficient by 6, yielding a null space containing at least six null solutions \vctr{d} (mechanisms), three of which will be the trivial translations \cite{Guest03}. The null eigenvectors are the modes of the framework that do not strain the O-O linkages. As the framework folds, the $\mat{e}=0$ condition holds to first order between each step, but $\mat{C}$ must be continually updated when finite mechanisms explore their range.


When there is no applied applied tension field (zero pressure), the equation of motion is
\begin{eqnarray}
m\ddot{\vctr{d}} & = & -k\mat{C}^{\rm T}\mat{C}\,\vctr{d}.
\label{eq:EOM2}
\end{eqnarray}
$m$ is the oxygen mass (we ignore Si atoms). We ignore Si-O-Si bending forces, so $k$ is the spring constant of the O-O `bars'. Solutions of the type $\mat{d}_{l}(t) = \mat{d}_{l}\exp(i\omega_{l}t)$ yield the eigenvalue equation
\begin{equation}
\mat{C}^{\rm T}\mat{C}\,\mat{d}_{l} = \frac{m}{k}\omega_{l}^{2}\,\mat{d}_{l} = \nu_{l}^{2}\,\mat{d}_{l},
\label{eq:ev}
\end{equation}
which is valid for small displacements. Removing the trivial translations, $\mat{C}^{\rm T}\mat{C}$ is a $(3N\!+\!3)\! \times\! (3N\!+\!3)$ square matrix. This underconstrained eigenvalue equation can be solved using singular value decomposition (SVD) \cite{Pellegrino86}, and the null modes, $\omega_{l}=0$, define the infinitesimal mechanisms $\mat{d}_{l}$ consistent with the flexibility window, $\mat{e}=0$.

Thermodynamically, heat provides the energy that allows the system to explore its allowed configurations. In the limit of high temperature $T$, the vibrational entropy of a system of uncorrelated oscillators is
\begin{equation}
S = -k_{\rm B}\sum_{l=1}^{3N+3}\log\left( \frac{\hbar \omega_{l}}{k_{\rm B}T} \right).
\label{eq:entropy}
\end{equation}
The product of eigenvalues equals the determinant, so
the vibrational entropy is then
\begin{equation}
S \!=\! -k_{\rm B}
\left[
\frac{1}{2}\log\left( \gamma^{3N+3}\det\left[ \mat{C}^{\rm T}\mat{C} \right]\right)
\right],
\label{eq:vibent}
\end{equation}
where $\gamma\!=\!(\hbar/k_{\rm B}T)^{2}k/m$.
This represents an entropic density, since (\ref{eq:vibent}) applies to each configuration $\mat{C}$ within the flexibility window. Although this expression includes contributions from vibrational modes, the $\omega_{l}\!=\!0$ null modes will dominate the entropic density since the logarithm diverges. To avoid divergences in (\ref{eq:vibent}), we added $10^{-5}$ to the null modes ($\nu_{l}$), rendering them `floppy'. For large-amplitude mechanisms, we do not expect (\ref{eq:vibent}) to be accurate. 

To explore the dependence on the size of the periodic unit, the unit cell can be expanded along the three axes to generate a supercell containing $N_{c}$ unit cells. The number of modes increases linearly with $N_{c}$. Depending on the framework, the folding mechanisms will grow as $N_{c}^{n/3}$ ($0 \le n \le 3$). When $n\!=\!3$ the entropic contribution from null modes is extensive, growing linearly with volume.
The computational effort generally increases with the cube of the volume explored, i.e.~$\sim N_{c}^{3}$. To speed up computation, we applied Bloch's theorem $\vctr{d}(\vctr{r}\!+\!\vctr{R}) \!=\! \vctr{d}(\vctr{r}) \exp(2\pi i \vctr{K}\! \cdot\! \vctr{R})$ to a single unit cell to explore the long wavelength null mode states within the Brillouin zone, and not just those at $\vctr{K}\!=\!0$ \cite{Hutchison06}. \vctr{R} is the position vector of a neighboring cell. The wavevector $\vctr{K}$ has components that are multiples of $(1/(pa), 1/(qb), 1/(rc))$, where $p$, $q$ and $r$ are the cell repeats along the $a$, $b$ and $c$ axes (thus~$N_{c}\!=\!pqr$). Because of loss of periodicity, Bloch's theorem can not be used to follow a mechanism to finite amplitude. We use it to identify mechanisms, but follow them as modes at the $\Gamma$ point ($\vctr{K}=0$) in the $p,q,r$-expanded cell.

Our computational tool, Zeolite Nullspace Explorer (ZeNuSpEx), follows the mechanisms as the structure evolves through the flexibility window. To initialize \mat{C}, geometrically-relaxed initial coordinates were obtained using the GASP program \cite{GASP}. GASP binds the oxygen atoms to the vertices of ideal tetrahedra by springs of natural length, and then minimizes the spring energy using the conjugate gradient method. \mat{C} was adjusted continually by applying second-order corrections as the structure evolves. Occasionally, \mat{C} is regularized by applying GASP. The singular value decomposition is the rate-determining step \cite{NumRecipes92}. We applied ZeNuSpEx to the 197 known zeolite frameworks and found that 25 of them exhibit extensive flexibility (Table \ref{tab:extflex}).
\begin{table}
\caption{\label{tab:extflex}The twenty-five extensively flexible zeolite frameworks. The number of mechanisms at a general reciprocal lattice point is given. The `*' indicates that the framework could not be relaxed with perfect tetrahedra.}
\begin{ruledtabular}
\begin{tabular}{l@{ -}ll@{ -}ll@{ -}ll@{ -}ll@{ -}l}
\IZA{ACO} & 4  &  \IZA{EMT} & 2  &  \IZA{LTA} & 4  & \IZA{OFF} & 1  &  \IZA{SSF} & 2 \\
\IZA{AST} & 12 &  \IZA{ERI} & 2  &  \IZA{LTL} & 3  & \IZA{PAU} & 24* &  \IZA{TSC} & 25 \\
\IZA{ASV} & 2  &  \IZA{FAU} & 16 &  \IZA{MER} & 2  & \IZA{RHO} & 4  &  \IZA{UFI} & 2 \\
\IZA{DFO} & 4  &  \IZA{KFI} & 6  &  \IZA{MOZ} & 2  & \IZA{SAS} & 2  &  \IZA{UOS} & 2 \\
\IZA{EAB} & 1  &  \IZA{LEV} & 3  &  \IZA{MTN} & 2* & \IZA{SOD} & 2  &  \IZA{UOZ} & 4 \\
\end{tabular}
\end{ruledtabular}
\end{table}

\begin{figure}[t!]
	\includegraphics[scale=0.60]{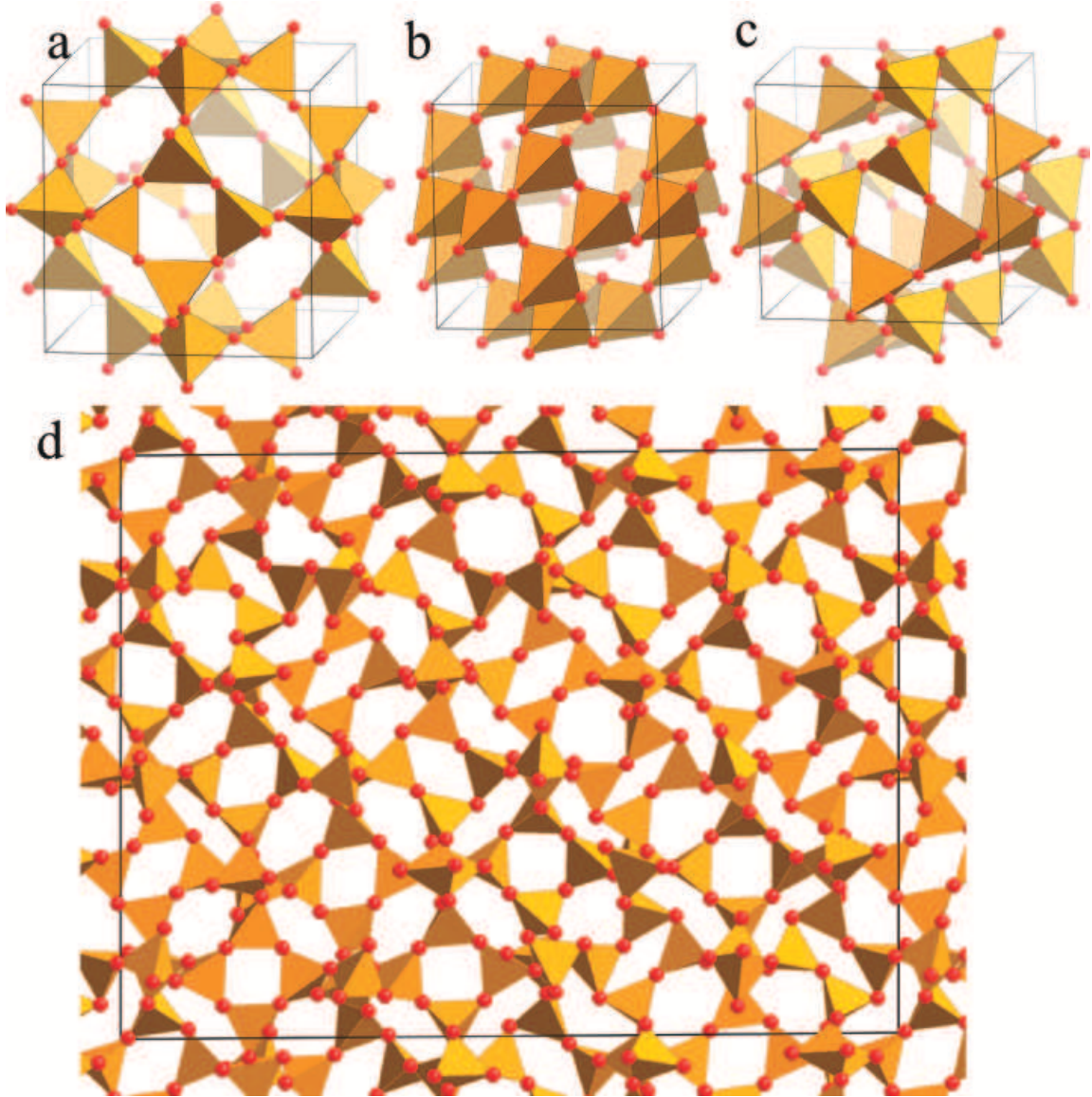}
	\caption{(a) \IZA{SOD} cage in its minimum density (and maximum symmetry $I m\overline{3}m$) flexible state. (b) A higher density folded state that preserves $I \overline{4}3m$ symmetry. (c) A folded state in $P1$ symmetry at its maximum density. (d) A system of $4\!\times\!3\!\times1$ unit cells of \IZA{SOD}, viewed down [001], showing a higher-order mechanism in $P1$ symmetry. The spheres are the centers of oxygen atoms.}
	\label{fig:SOD}
\end{figure}

As an example, Fig.~\ref{fig:SOD}a shows the \IZA{SOD} framework in its maximum symmetry ($Im\overline{3}m$), minimum density state, 16.58 T atoms/nm$^{3}$. Figs.~\ref{fig:SOD}b and \ref{fig:SOD}c show the framework folded along two $\vctr{K}=0$ mechanisms to their densest configurations, 27.98 and 22.54 T atoms/nm$^{3}$, respectively. The minimum density is determined by the onset of stretching of O-O edges; the maximum density is determined by oxygen overlaps, assuming an oxygen radius of $L/2\!=\! 0.135$ nm. The body-centered unit cell of \IZA{SOD} supports 10 finite internal mechanisms at $\vctr{K}=0$. This agrees with Hammonds et al.~if we add their $\Gamma$ and $H$ modes for the primitive cell \cite{Hammonds98}. The minimum density state (Fig.~\ref{fig:SOD}a) represents a confluence for all mechanisms, and so the density of states for null modes is highest at this end of the flexibility window. Few modes reach the densest states.


As we increase the supercell size ($\vctr{K}\ne 0$) so that it contains $N_{c}$ primitive cells, the number of mechanisms increases linearly with system volume. A mechanism for a $4\times3\times 1$ supercell of \IZA{SOD} is depicted in Fig.~\ref{fig:SOD}d. All vibration modes and mechanisms can be excited simultaneously at small amplitude at the maximum unit cell volume condition. Once a mechanism is excited at large amplitude, lowering the symmetry and increasing the framework density,  mechanisms that are incompatible with the lowered symmetry are suppressed.

Monte Carlo simulations of random mechanism excitations in \IZA{SOD} allow the system to explore the flexibility window. With increasing $N_{c}$, the system spends significantly more time near the minimum density state since fewer random moves will correspond to cooperative folding modes; the system spends most of its time where the density of mechanisms is highest.

\begin{figure}[t!]
	\includegraphics[scale=0.8]{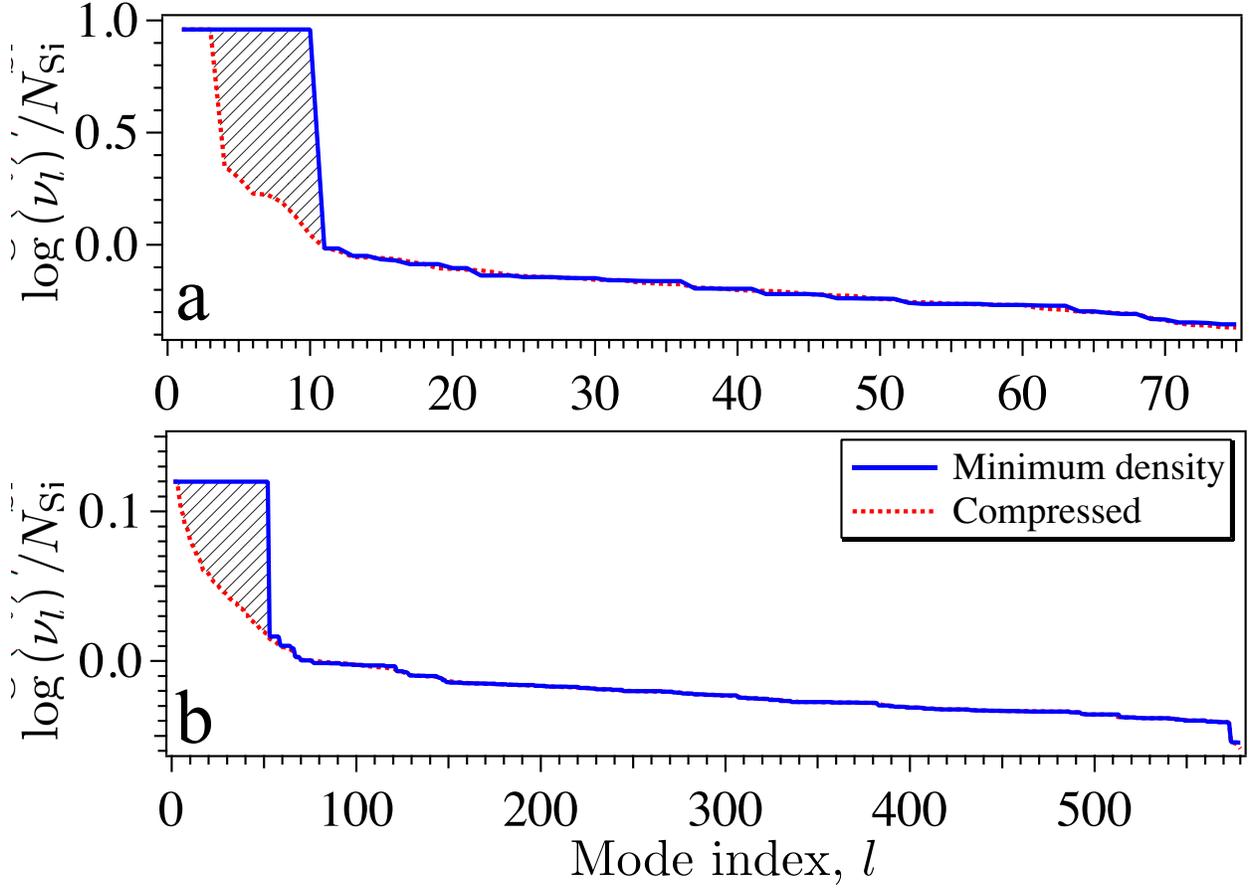}
	\caption{Plots of $\log(\nu_{l})$ per \tetra\ tetrahedron for the \IZA{SOD} framework, sorted by increasing frequency, for the minimum density configuration (solid lines) and a randomly-selected folded state (dashed lines). (a) Single unit cell ($\vctr{K}\!=\!0$ modes). There are 75 modes, 10 of which are null modes at maximum symmetry, $I\overline{3}m$, reducing to 3 in the folded $P1$ symmetry. (b) $2 \!\times\! 2\! \times\! 2$ supercell of the \IZA{SOD} framework. 579 modes, 52 null modes, reducing to 3 in the folded state.}
	\label{fig:states}
\end{figure}

The contribution to the configurational entropy can be assessed by plotting $\log \left( \nu_{i} \right)/N_{\rm Si}$ against the mode index, sorted in order of increasing frequency (Fig.~\ref{fig:states}). In Fig.~\ref{fig:states}a we show the $\vctr{K}\!=\!0$ mode distribution for the minimum density and a folded configuration in $P1$ symmetry. The folded structure, having $P1$ symmetry, retains just 3 null modes. The shaded region represents the loss in configurational entropy when the structure folds. The equivalent plot for a $2\!\times\!2\!\times\!2$ unit cell repeat (Fig.~\ref{fig:states}b) shows that the total number of modes increases to 52. There is a decrease in the shaded area, but with further increases in system size the mode distribution remains similar to that shown in Fig.~\ref{fig:states}b. In the asymptotic limit of large $N_{c}$, the count of mechanisms grows extensively, two mechanisms per unit cell.

Extensive flexibility is found to be associated with structures that have high symmetry. Conversely, $P1$ guarantees just three infinitesimal modes, regardless of system size. Frameworks with higher topological symmetry always exhibit the highest density of mechanisms at the low density end of the flexibility window. A topologically high-symmetry structure, once folded to $P1$, is guaranteed only three mechanisms, of which at least one must be finite since this folded state is accessible from the high symmetry state.

\begin{figure}[t!]
	\includegraphics[scale=0.7]{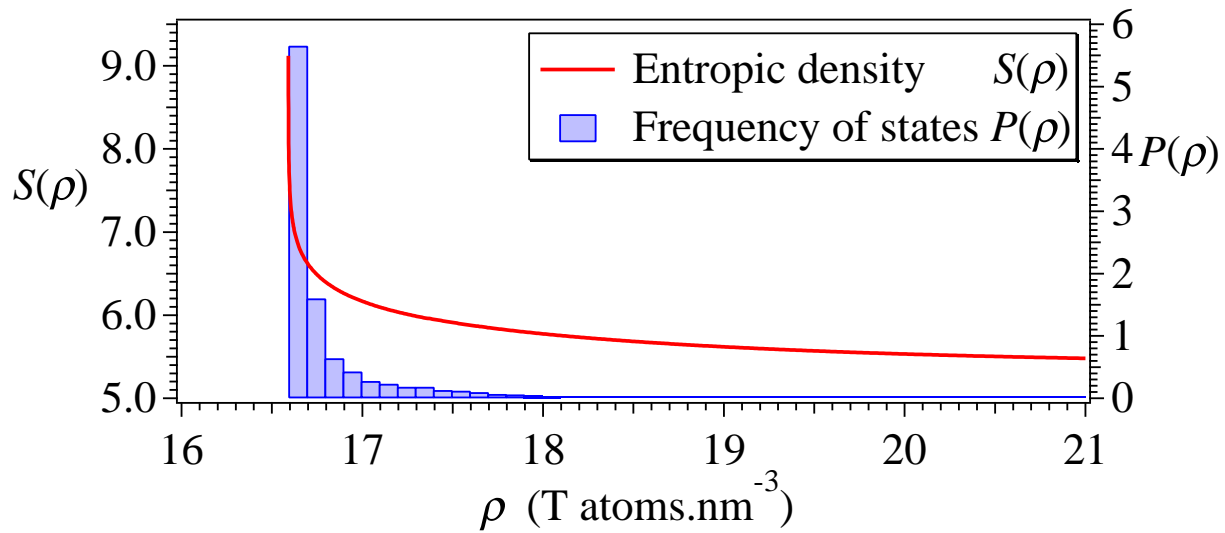}
	\caption{Plot of entropic density $S(\rho)$ as a function of $\rho$. $S(\rho)$ is sharply peaked at the minimum density $\rho = 16.61$ T atoms/nm$^{3}$, where the framework symmetry is maximum. All mechanisms pass through this state. Superimposed on the plot is the frequency of states $P(\rho)$) arising when 7000 random-walk paths are followed throughout the flexibility window. The flexibility window has one mode that reaches the highest density of 30.29 T atoms/nm$^{3}$. The random paths never reached this state, but frequently encountered the minimum density state.}
	\label{fig:EntDensity}
\end{figure}

As a rule, the minimum density configuration is also the maximum symmetry state. The highest symmetry generally has the largest degeneracy of constraints, and thus the excess of degrees of freedom over constraints is maximum. Every mechanism supported by the structure has a path that passes through the minimum density state, which therefore represents the confluence of all mechanisms. Mechanisms can be linearly superimposed, with infinitesimal amplitude, at this lowest density state. Not all mechanisms are capable of reaching the high density end of the flexibility window, and so there is a gradient of null states that peaks at minimum density, and usually decreases monotonically as framework density increases. Fig.~\ref{fig:EntDensity} shows how the entropic density term, $S(\rho)$, depends on framework density, $\rho$, near the low-density end of the flexibility window. The curve is the average over 7000 different paths through the window. Superimposed is the average number of mechanisms available $P(\rho)$ at each density. Both curves show a smooth decrease away from the minimum density state. Earlier, Duxbury et al.~\cite{Duxbury99} reasoned that the negative of the number of floppy modes in Bethe lattices behaves as a free energy. As we show, in zeolites extensive floppy modes reduce free energy by increasing the configurational entropy.

Almost all of the 197 known zeolite frameworks have a flexibility window, whereas 12\% of them are extensively flexible (see Table~\ref{tab:extflex}). It appears that the existence of a flexibility window alone is important, and that the extensivity of flexible modes, despite an apparent thermodynamic advantage of increased configurational entropy in bulk crystals, is not essential. As pointed out by Hammonds et al.~\cite{Hammonds97} local flexible deformations can accommodate cations, or structure-directing molecules. Similarly, the surface and edge mechanisms of the flexibility window, which grow non-extensively as $N_{c}^{2/3}$ and $N_{c}^{1/3}$ respectively, may play a crucial role at the nucleation and growth stage of zeolite synthesis where $N_{c}$ is small. In our database of hypothetical zeolites, only about 10\% of the low framework energy frameworks exhibit a flexibility window. It is clear that a flexibility window is the hallmark of a viable zeolite framework, and provides an important selection criterion in addition to low framework energy when evaluating databases for realizable structures.


We are grateful for financial support from the National Science Foundation grants Nos.~DMR-0703973 and DMS-0714953. Rivin thanks the Institute for Advanced Study and the Insitut f\"ur Mathematik, Techniche Universit\"at Berlin for support

\newpage

\newpage

\newpage

\end{document}